\documentclass[letter,10pt,twocolumn]{book}
\usepackage{epsfig}
\usepackage{taylor01}
\def\q{\qquad}
\def\beg{\begin{eqnarray}}
\def\ende{\end{eqnarray}}

\def\Rey{\mbox{\rm Re}}   
\def\Om{\it \Omega}
   
\begin{document}
\twocolumn[%
    \Talk         
\centerline{\it 12$^{th}$ International Couette-Taylor Workshop, September 6-8,
2001, Evanston, IL USA}
\Title{MHD instability in cylindric Taylor-Couette flow}

\Authors{
\Author[gruediger@aip.de]{1}{G. }{R\"udiger} and
\CoAuthor{1}{D.A.}{Shalybkov}
}
\begin{Adresses}
\Institute{1}{Astrophysikalisches Institut Potsdam}{ Germany}
\end{Adresses}
\Abstract{ The linear marginal instability of an  axisymmetric   MHD Taylor-Couette flow of 
infinite vertical extension is considered. 
For flows with a resting outer cylinder there is a well-known characteristic Reynolds number 
even without magnetic field but for sufficiently weak magnetic fields there are solutions 
with {\em smaller} Reynolds numbers so that a characteristic minimum exists.     The minimum 
only exists, however,  for not too small magnetic Prandtl numbers. 
 For small magnetic Prandtl numbers one only finds the typical magnetic suppression of the instability.

We are  here particularly interested in the case where  
the outer cylinder rotates so fast that the Rayleigh criterion for 
{\em hydrodynamic} stability is fulfilled.  We find that for given magnetic Prandtl number now 
always a magnetic field amplitude exists where the characteristic Reynolds number is minimal.  
These critical values are  computed for different magnetic Prandtl numbers. In all cases 
the Reynolds numbers are running with 1/Pm so that for the  small magnetic Prandtl numbers 
of sodium (10$^{-5}$) or gallium (10$^{-6}$) the critical Reynolds numbers 
exceed values of 10$^6$ or 10$^7$, resp. 
}

]

\subsection*{Introduction}
The longstanding problem of the generation of turbulence in various
hydrodynamically stable situations has found a solution in recent years
with the so called `Balbus-Hawley instability', in
which the presence of a magnetic field has a destabilizing effect on a
differentially rotating flow, provided that the angular velocity decreases
outwards with the radius.
This magnetorotational instability (MRI)  has been  discovered decades ago  for
ideal Couette flow, but it  has never been observed in the laboratory. Moreover, Chandrasekhar (1961) already suggested  the existence of MRI for ideal Taylor-Couette flow, but
his results for non-ideal fluids for small gaps and within the  small magnetic Prandtl
number approximation  demonstrated the absence of MRI for hydrodynamically stable and/or 
unstable flow. 
Recently, Goodman and Ji  (2001) claimed that this absence of MRI
was due to the use of the small magnetic Prandtl number limit. 
The magnetic Prandtl number Pm$= \nu/\eta$ is really very small under laboratory conditions.  Obviously, the understanding of this phenomenon  is very
important for possible experiments, Taylor-Couette flow  dynamo experiments included.

Here  the dependence of the magnetic Prandtl number on the MHD instability of 
Taylor-Couette flow is investigated. The simple model of uniform  density fluid
contained between two vertically-infinite rotating cylinders is used with constant magnetic
field parallel to the rotation axis. 
For viscous flows then the most general form of $\Om$ is
\begin{equation}
\Om(r) = a+b/{R}^2,
\label{Om}
\end{equation}
where $a$ and $b$ are two constants related to the angular
velocities $\Om_{\rm in}$ and $\Om_{\rm out}$ with which the inner
and the outer cylinders are rotating. If $R_{\rm in}$ and $R_{\rm out}$
($R_{\rm out}>R_{\rm in}$) are the radii of the two cylinders then
\begin{equation}
a=\Om_{\rm in}{\hat \eta}^2{\hat \mu-{\hat\eta}^2\over1-{\eta}^2}
\; \; {\rm and} \; \;
b=\Om_{\rm in} R_{\rm in}^2 {(1-\hat\mu)\over1-{\hat\eta}^2},
\label{ab}
\end{equation}
with 
\begin{equation}
\hat\mu=\Om_{\rm out}/\Om_{\rm in}  \q  {\rm and} \q 
\hat\eta=R_{\rm in}/R_{\rm out}.
\label{mu}
\end{equation}
After the Rayleigh stability criterion, $d(R^2 \Om)/dR>0$, rotation laws with positive numbers $a$ are 
hydrodynamically stable, i.e. for $\hat\mu>\hat\eta^2$. Taylor-Couette flows with resting 
outer cylinders ($\hat\mu=0$) are thus never stable.

\begin{figure}
\psfig{figure=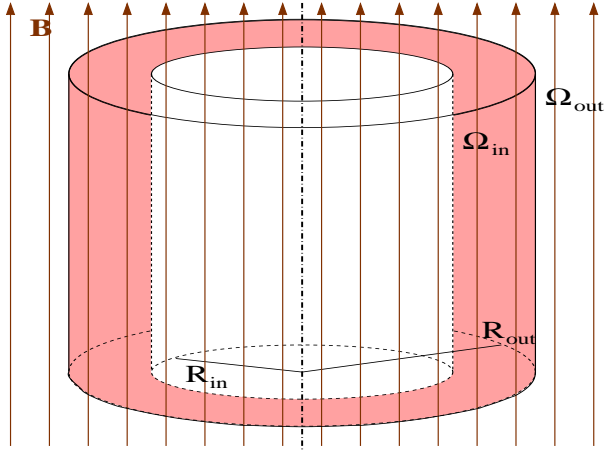,width=8cm,height=6cm}
\caption{Cylinder geometry of the  Taylor-Couette flow}
\label{geometry}
\end{figure}

\subsection*{Boundary conditions}
An appropriate set of ten boundary conditions is needed to solve  
the equation system
(see R\"udiger and Zhang 2001).  Always no-slip conditions for the velocity on the walls 
are used, i.e. 
\beg
u_R=0, \q u_\phi=0, \q {d u_R \over dR}=0.
\label{bvel}  
\ende
The magnetic boundary conditions depend on the electrical properties
of the walls. The transverse currents and perpendicular component of
magnetic field should vanish on conducting walls, hence
\beg
{d b_\phi \over dR} + {b_\phi \over R}=0, \q b_R=0.
\label{cond}
\ende
The above boundary conditions (\ref{bvel}) and (\ref{cond}) are valid 
for $R=R_{\rm in}$ and  for $R=R_{\rm out}$. 

Rotation and magnetic field are represented in the numerical simulations by the 
Reynolds number Re$= \Om_{\rm in} R_{\rm in} (R_{\rm out} - R_{\rm in})/\nu$ 
and the Hartmann number Ha$=(R_{\rm in}(R_{\rm out} - R_{\rm in})/\mu_0 \rho \nu 
\eta)^{1/2} B$.

\subsection*{Resting outer cylinder}
\begin{figure}
\psfig{figure=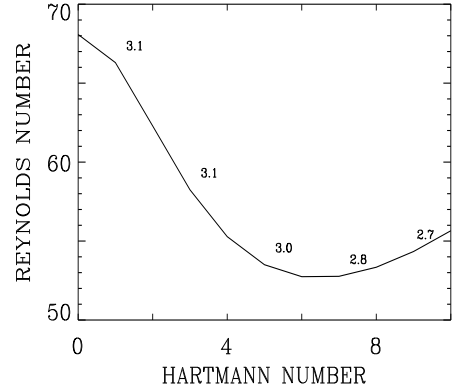,width=7.0cm,
height=6cm} 
\caption{The Reynolds numbers for Taylor-Couette flow with resting outer 
     cylinder with $\hat\eta=0.5$ and  for  Pm=1. There is instability even without magnetic 
     fields but  its excitation is much easier with magnetic  fields with 
     Hartmann numbers of about 6...7. The  line is marked with those  
     wavenumbers for which the eigenvalues are minimal. } 
\label{fig}
\end{figure}

\begin{figure}
\psfig{figure=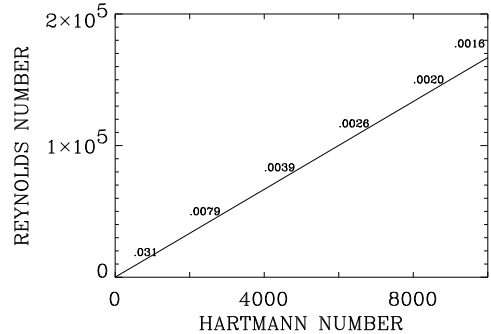,width=7.0cm,
height=5cm} 
\caption{The  same as in Fig. \ref{fig} but   for  Pm=$10^{-5}$.  The minimum characteristic for Pm=1 completely disappears.} 
\label{fig0}
\end{figure}

In Fig. \ref{fig}  a resting outer 
cylinder is considered for a medium-size gap of $\hat\eta$=0.5 and for Pm=1. As 
we know for vanishing magnetic field the exact 
Reynolds number for this case is  about 68 -- well represented by the result for Ha=0  in Fig. \ref{fig}. 
But for increasing magnetic field  the Reynolds number is reduced so that the excitation of the 
Taylor vortices 
becomes easier 
than without magnetic field. The minimum Reynolds number Re$_{\rm crit}$ of about 52.5 for Pm=1
is reached for Ha$_{\rm crit}\simeq$ 6...7.  This  magnetic induced subcritical excitation of Taylor vortices is due to the MRI.
Always for a (say) critical Hartmann number the Reynolds numbers 
take a minimum which we shall call the critical Reynolds numbers.  For even stronger 
magnetic fields -- as it must be -- the magnetic field starts to suppress the 
instability  
(see also R\"udiger and Zhang 2001).
In Fig. \ref{fig0}  the same container is considered but for the small magnetic 
 Prandtl number of $10^{-5}$. The minimum  characteristics for Pm=1 completely disappears, only  suppression of the instability by the magnetic field can be observed.

\subsection*{Rotating outer cylinder}
\begin{figure}
\psfig{figure=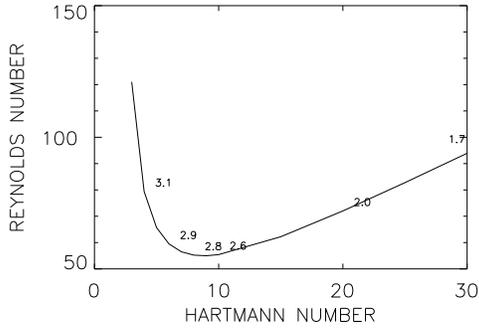,width=7.0cm,
height=5cm}
     \caption{MHD instability map for  Taylor-Couette flow for $\hat\eta=0.5$ and  Pm=1. The outer cylinder  rotates with 33\% of the rotation rate of the inner cylinder so that after the Rayleigh criterion the hydrodynamic instability disappears. Again the  curve is marked with the critical wavenumbers} 
\label{fig2}
\end{figure}

Another situation holds if  the outer cylinder rotates so fast that the rotation law 
does no longer fulfill the Rayleigh criterion and a solution for Ha=0  cannot 
    exist. Then the nonmagnetic eigenvalue along the vertical axis moves to 
    infinity and we should always have a minimum.   It is the basic situation 
    in astrophysical applications such for accretion disks with a Kepler 
    rotation law. Here the question is whether the critical Reynolds number and the critical Hartmann number can experimentally be realized. 
The Figs. \ref{fig2}...\ref{fig4}  present the results for both various Hartmann numbers and magnetic Prandtl 
numbers for a medium-sized gap of $\hat\eta$=0.5. 

\begin{figure}
\psfig{figure=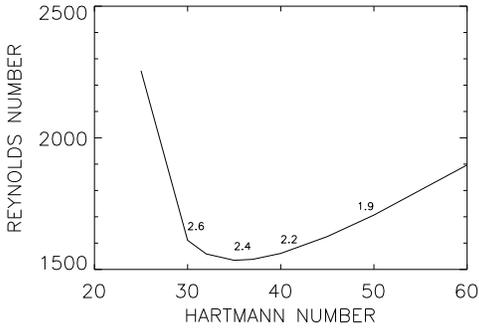,width=7.0cm,
height=5cm} 
     \caption{The same as in Fig. \ref{fig2} but for  Pm=0.01} 
\label{fig3}
\end{figure}

\begin{figure}
\psfig{figure=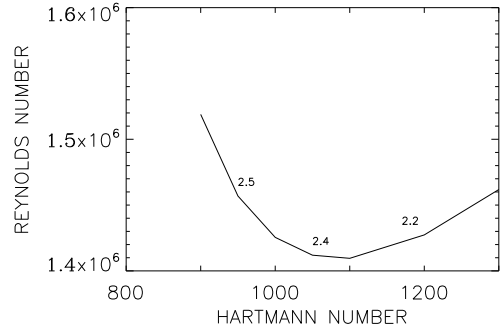,width=7.0cm,
height=5cm} 
     \caption{The same as in Fig. \ref{fig2} but for Pm=$10^{-5}$} 
\label{fig4}
\end{figure}
\begin{figure}
\psfig{figure=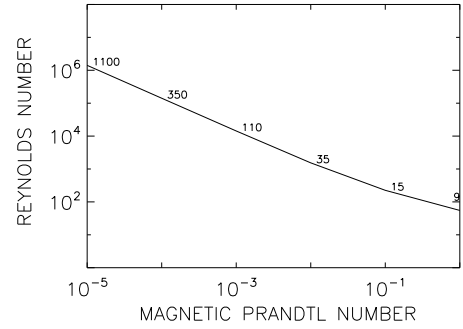,width=7.0cm,
height=5cm}
     \caption{The critical Reynolds numbers for given Prandtl numbers  marked 
with those Hartmann numbers where the Reynolds number is minimal.} 
\label{fig5}
\end{figure}

There are always minima of the characteristic Reynolds numbers for  certain Hartmann numbers. The minima and the critical Hartmann numbers increase for decreasing magnetic Prandtl numbers.
For $\hat\eta$=0.5 and $\hat\mu$=0.33 the  critical Reynolds numbers together with the critical Hartmann 
numbers are plotted in Fig. \ref{fig5}. 

For  the small magnetic Prandtl numbers we find interesting and simple relations. With
\beg
C_\Omega = {\rm Re} {\rm Pm}
\label{COm}
\ende
and
\beg
{\rm Ha}^* = {\rm Ha} \sqrt{\rm Pm}
\label{Ha-st}
\ende
it follows
\beg
C_\Omega \simeq 14
\label{COmega}
\ende
and
\beg
{\rm Ha}^* \simeq 3.3.
\label{HA}
\ende
$C_\Omega$ is the magnetic Reynolds number, $C_\Omega = \Om_{\rm in} 
R_{\rm in}(R_{\rm out} - R_{\rm in})/\eta$ 
(or dynamo number) and Ha$^*$ is the magnetic Hartmann number Ha$^* = 
(R_{\rm in}(R_{\rm out}- R_{\rm in})/\mu_0 \rho \eta^2)^{1/2} B$.

\
\end{document}